\documentclass[a4paper,fleqn,usenatbib]{mnras}

\usepackage[T1]{fontenc}
\usepackage{ae,aecompl}

\usepackage{longtable}
\usepackage{natbib}
\usepackage{psfrag}
\usepackage{xspace}
\usepackage{xcolor}
\usepackage{graphicx}   % Including figure files
\usepackage{amsmath}    % Advanced maths commands
\usepackage{amssymb}    % Extra maths symbols
\usepackage[normalem]{ulem}

\newcommand{\dik}{\textcolor[rgb]{0.,0.,0.}}%{0.0,0.69,0.31}}

\title[The planetary atmosphere evolution in TOI-942 and TOI-421 systems]{Atmospheric mass loss and stellar wind effects in young and old systems \dik{II: Is TOI-942 the past of TOI-421 system?}}

\author[D. Kubyshkina et al.]{
Daria Kubyshkina$^{1,2}$\thanks{E-mail: kubyshkd@tcd.ie}, 
Aline A.~Vidotto$^{1,3}$, 
Carolina Villarreal D’Angelo$^{4}$, \newauthor \hspace{0.08 cm}
Stephen Carolan$^{1}$, 
Gopal Hazra$^{1,3}$,
Ilaria Carleo$^{5,6}$
\\
$^{1}$School of Physics, Trinity College Dublin, the University of Dublin, College Green, Dublin-2, Ireland\\
$^{2}$Space Research Institute, Austrian Academy of Sciences, Schmiedlstrasse 6, A-8042 Graz, Austria\\
$^{3}$Leiden Observatory, Leiden University, PO Box 9513, 2300 RA, Leiden, The Netherlands\\
$^{4}$ Instituto de Astronomía Téorica y Experimental (CONICEt-UNC). Laprida 854, X5000BGR. Córdoba, Argentina\\
$^{5}$Astronomy Department and Van Vleck Observatory, Wesleyan University, Middletown, CT 06459, USA\\
$^{6}$ INAF – Osservatorio Astronomico di Padova, Vicolo dell’Osservatorio 5, I-35122, Padova, Italy\\
}

\date{Accepted XXX. Received YYY; in original form ZZZ}

\pubyear{2021}

\begin{document}
\label{firstpage}
\pagerange{\pageref{firstpage}--\pageref{lastpage}} \maketitle

\begin{abstract}
%241 of 250 words limit I SHORTENED IT A BIT TO FIT THE LIMIT
\dik{The two planetary systems TOI-942 and TOI-421 share many similar characteristics, apart from their ages (50~Myr and 9~Gyr). Each of the stars hosts two sub-Neptune-like planets at similar orbits and in similar mass ranges. In this paper, we aim to investigate whether the similarity of the host stars and configuration of the planetary systems can be taken as proof that the two systems were formed and evolved in a similar way. 
In paper I of this series, we performed a comparative study of these two systems using 3D modeling of atmospheric escape and its interaction with the stellar wind, for the four planets. We demonstrated that though the strong wind of the young star has a crucial effect on observable signatures, its effect on the atmospheric mass loss is minor in the evolutionary context. Here, we use atmosphere evolution models to track the evolution of planets in the younger system TOI-942 and also to constrain the past of the TOI-421 system. We demonstrate that despite all the similarities, the two planetary systems are on two very different evolutionary pathways. The inner planet in the younger system, TOI-942, will likely lose all of its atmosphere and become a super-Earth-like planet, while the outer planet will become a typical sub-Neptune. Concerning the older system, TOI-421, our evolution modeling suggests that they must have started their evolution with very substantial envelopes, which can be a hint of formation beyond the snow line.}
\end{abstract}


%%%%%%%%%%%%%%%%%%%%%%%%%%%%%%%%%%%%%%%%%%%%%%%%%%

%%%%%%%%%%%%%%%%% BODY OF PAPER %%%%%%%%%%%%%%%%%%

\section{Introduction}\label{sec::intro}

\dik{Each individual planetary system and the observed population of exoplanets in general were shaped by a combination of planetary formation processes, including the various scenarios of planetary core formation, atmospheric accretion, and planetary migration, as well as atmospheric mass loss \citep[see, e.g.,][]{fulton2017,owen_wu2017,jin_mordasini2018,gupta_schlichting2019,gupta_schlichting2020,loyd2020,sandoval2021}. However, the relative role of these processes is not precisely known and is currently widely debated. Even when atmospheric mass loss models can be constrained by observations, an initial state of individual planetary systems and the formation processes that led to them remains unknown. Fortunately, it has been demonstrated that coupling atmospheric evolution models and present-day observations of exoplanets can provide constraints on the initial parameters of planetary systems and the early stages of evolution of their host stars \citep{kubyshkina2019a,kubyshkina2019b,owen2020entropy,rogers_owen2021,bonfanti2021}. 
Of particular interest is applying such methods to multiplanetary systems, as they allow to better constrain the early properties of the system and some of the poorly known present-day parameters \citep{kubyshkina2019a,kubyshkina2019b,owen_estr2020}. Additionally, applying these coupled models to different systems with similar architecture (i.e., comparative studies) allows us to constrain whether these similar systems underwent similar formation scenarios. This is precisely what we investigate here.}

\dik{In the present study, we model the evolution of planetary atmospheres in two systems of very similar configuration, but very different ages, the 9-Gyr old TOI-421 \citep{Carleo_toi421} and the 50-Myr old TOI-942 \citep{Carleo_toi942}. 
These two host-stars have similar masses of about $0.85$~$M_{\odot}$ and close spectral types (G9 and K2.5V) and each of them hosts two sub-Neptune-like planets at similar close-in orbits about $0.05$ and $0.1$~AU, where the evolution of their atmospheres is to a large extent controlled by atmospheric mass loss driven by stellar irradiation. The masses of companion planets also lay in similar ranges: for planets b and c, $\sim$7.2~$M_{\oplus}$ and 16.4~$M_{\oplus}$ in the older system TOI-421, and below 16~$M_{\oplus}$ and 37~$M_{\oplus}$ in the younger system TOI-942.} %All four planets are expected to undergo strong atmospheric escape.

\dik{In the first paper of this series \citep[][further referred to as Paper I]{kubyshkina2021tois1}, we performed 3D atmospheric escape modeling to study in detail the interaction between the evaporating planetary atmospheres and the stellar winds. We found that, though the stellar winds can affect the observable signatures of escaping atmospheres significantly, they have only a minor effect on the atmospheric mass loss rates of these planets, and that this is negligible in the evolutionary context. Therefore, we concluded in Paper I that using evolutionary models that do not account for the stellar wind effects can be safely applied to study these systems.}

\dik{To track the evolution of planetary atmospheres, here we employ the framework presented in \citet{kubyshkina2020mesa,kubyshkina2021mesa} that couples the thermal evolution of a planet with a realistic prescription of the atmospheric mass loss based on hydrodynamic modeling. Here, we further update this framework by using the extended grid of planetary upper atmospheres models presented in \citet{kubyshkina2021RN}. Our work explores two approaches: a forward modelling, in which the young system TOI-942 is evolved in time up to the age of the old system, and a backward modelling, in which the old system TOI-421 is `rejuvenated' (evolved back in time) down to the age of the young system. The second modeling approach, as we will show, is more costly and less trivial.}

\dik{We expect the planets in the TOI-942 system (50-Myr old) to be at the end of an extreme atmospheric escape phase. This is because atmospheric mass loss is particularly intense within the first few tens of Myr, when the planet is hot and inflated and is subjected to the strong high-energy irradiation of the young host star \citep[e.g.,][]{watson1981,lammer2003,owen_wu2016,fossati2017,kubyshkina2018_k2-33b}. This is consistent with the relatively high atmospheric mass-loss rates obtained with our 3D modeling in Paper I: 5.1-12$\times10^{11}$~${\rm g~s^{-1}}$ for the inner planet, and 3.2-3.5$\times10^{11}$~${\rm g~s^{-1}}$ for the outer one. These values are further confirmed by our evolutionary models here and suggest that we can make a robust prediction of the future of this system.}

%\dik{We demonstrate that the young planets b and c in the TOI-942 system are unlikely to follow the same evolutionary scenario as the planets b and c in the TOI-421 system, despite both b and c planets in the two systems being at similar close-in orbits and in similar mass ranges. Instead, the planetary system TOI-942 will presumably end up as one of the systems hosting two planets of relatively similar masses but lying on different sides of the radius gap, i.e., with a closer-in super-Earth and a further-out sub-Neptune planet. Examples of such systems are, e.g., HD~3167 \citep{christiansen2017} and Kepler-36 \citep{Vissapragada2020}. } %i am unsure if you need this paragraph in the introduction. it seems to be more adequate for a summary and conclusions session --> after reading the conclusions, i moved it to the conclusions, if you agree with that, then please remove it from here %Commented out, probably you're right

\dik{For the older system TOI-421, we face the opposite task and have to resolve the initial parameters of the system. To achieve this, we produce a large number of evolutionary models assuming different initial parameters for the system, including different initial planetary atmospheres and masses, and different activity levels of the host star, and check a posteriori if each of these tracks can fit the present-day observations. } 
%We find that planets in the TOI-421 system started their evolution with more massive envelopes compared to planets in the TOI-942 system, potentially suggesting their formation beyond the snow line. Another outcome of our models is that we can infer the rotation rate of TOI-421 at its youth. We show here that TOI-421 had likely a slower rotation when it had the same age as TOI-942.%same comment here relating to the second part of this paragraph: could be removed from the intro and moved to end of the paper --> i moved it to the conclusions, if you agree with that, then please remove the second part of this paragraph
%I am not sure if we can really write that we 'constrain' the early time rotation, our constraint is based on that the initial parameters resolved for the slow rotator are less ridiculous, it's not very strict

%The natural question arising when thinking of the similarities between the two planetary systems is if the older system TOI-421 could look in the past similar to the young TOI-942 system. Even though the relation of the radii of the outer planets in the two systems does not favor this scenario (as we discussed in Section~\ref{sec::target_systems}), we do not yet exclude this possibility, as the radius of the older planet, TOI-421~c, can be inflated due to the increasing bolometric luminosity of its host star (typical for the ages older than $\sim6$~Gyr, see Figure~\ref{fig::stellar_models}). 

\dik{The paper is organized as follows. In Section~\ref{sec::target_systems} we discuss the properties of the stellar hosts TOI-421 and TOI-942 and their planets and the possible activity/rotation evolution of TOI-421. In Section~\ref{sec::model}, we give a brief overview of our model, and in Section~\ref{sec::evolution} we discuss the possible evolutionary scenarios for both systems. We summarize our conclusions in Section~\ref{sec::conclusions}.}

%%%%%%%%%%%%%%%%%%%%%%%%%%%%%%%%%%%%%%%%%%%%
\section{Target systems and stellar evolution models}\label{sec::target_systems}
%
%\subsection{Stars}\label{sec::target_systems_star}
%We consider two systems around G(K?)-type stars TOI-421 and TOI-942.

\dik{The two target systems, TOI-421 and TOI-942, were discussed in detail in Paper I, and their detailed parameters can be found in Table~1 of that paper. Here, we present a brief overview of the relevant parameters.}

\dik{The two stars have masses coinciding within their observational uncertainties: $0.88\pm0.04$~$M_{\odot}$ (TOI-942) and $0.85^{+0.029}_{-0.021}$~$M_{\odot}$ (TOI-421), and belong to neighboring spectral types, with parameters of each star being close to the border between the two classes. In terms of activity, these stars have parameters typical for their ages ($50^{+30}_{-20}$~Myr for TOI-942 and $9.4^{+2.4}_{-3.1}$~Gyr for TOI-421), with TOI-942 being more active than average for stars of similar masses, but not an outlier (see Paper I). At the age of TOI-421, the rotation period and high-energy radiation of main-sequence stars lie in a narrow range for the specific stellar masses \citep{matt2015,Johnstone2020mors}. Altogether, this suggests that both stars can potentially be at different evolutionary stages of the same star.}

\begin{figure*}
\includegraphics[width=0.8\hsize]{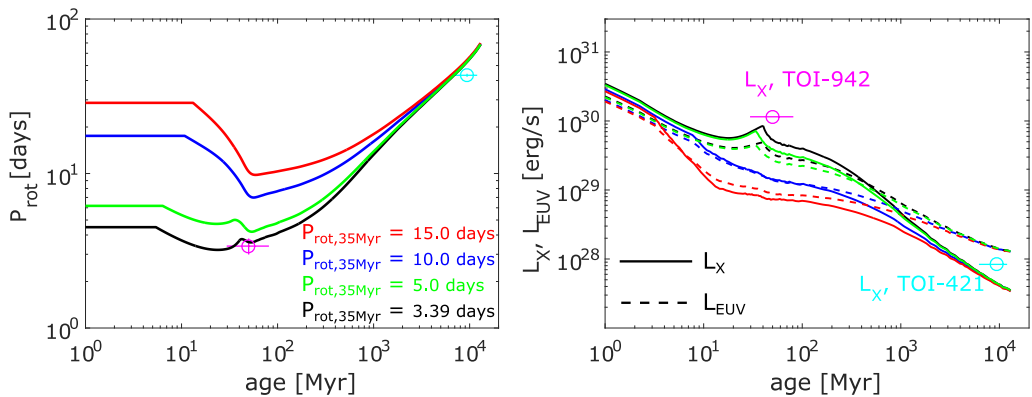}\\
\caption{\dik{Evolutionary tracks of TOI-421 and TOI-942 fitted using the Mors stellar evolution code \citep{Johnstone2020mors}. Here we show the {evolution of the} rotation period (left panel) and X-ray/EUV luminosities (right panel). {The black lines show the model corresponding to the stellar mass of 0.84~$M_{\odot}$, and a stellar rotation period of 3.39 days at the age of 35~Myr (``track 1'' in Paper I, and a fast rotator in this study). The green, blue, and red lines show the tracks assuming the same internal stellar parameters, but different rotation periods at 35~Myr: 5, 10, and 15 days, respectively.} The magenta/cyan circles show the present values for TOI-942/TOI-421 stars, as indicated in the right panel. The error bars for $P_{\rm rot}$ and the ages of the two systems are shown with vertical/horizontal lines. The given values of $L_{\rm X}$ are model dependent and are given for the median values.} }\label{fig::stellar_models}
\end{figure*}

\dik{In Paper I, we found 
%Applying the fitting procedure described in detail in Paper I, we use the Mors stellar evolution code \citet[][]{Johnstone2020mors,spada2013} to find 
the stellar evolutionary track that reproduces well the present-day parameters of TOI-942 and TOI-421. %This track corresponds to the assumption that the mass of TOI-942 is at its lower uncertainty border equal to 0.84~$M_{\odot}$, and the age of the star is 35~Myr. 
In this study, we will refer to this model as a fast rotator. In Figure~\ref{fig::stellar_models}, the black lines show the corresponding evolution of the rotation period ($P_{\rm rot}$, left panel) and of the X-ray and extreme ultraviolet luminosities ($L_{\rm X}$ and $L_{\rm EUV}$, respectively, right panel), that control to a large extent the atmospheric mass loss from the companion planets. The present-day values of $P_{\rm rot}$ and $L_{\rm X}$ are shown by red (TOI-942) and blue (TOI-421) circles and the measurements of $L_{\rm EUV}$ are not available.}

%\aav{add the initial period as annotation to left panel. for the right panel, maybe the max value in the y axis could be larger and the min(y) could also be higher: 3e28 to 1e31? }>>done

\dik{The fast rotator track from Figure~\ref{fig::stellar_models} is used to study the evolution of the TOI-942 system. However, as the initial rotation period of a star can be widely different \citep[see, e.g.,][]{pizzolato2003, mamajek2008, wright2011,jackson2012,tu2015, magaudda2020,Johnstone2020mors}, we cannot assume that TOI-421 has followed the same evolutionary path. We consider, therefore, three additional tracks, assuming different rotation periods at the age of 35~Myr: 5, 10, and 15 days (green, blue, and red lines, respectively). These periods correspond to moderate, slow, and very slow rotators according to the observations in the young stellar cluster NGC 2547 \citep{irwin2008}.}

\dik{Both TOI-942 and TOI-421 host two sub-Neptune-like planets in similar close-in orbits. In the TOI-942 system, planet b orbits at  $0.0498\pm0.0007$~AU and planet c at 
$0.088\pm0.0014$~AU, while in the TOI-421 system the orbits are $0.056\pm0.0018$~AU and $0.1189\pm0.0039$~AU, for planets b and c, respectively. Planetary masses in the older system TOI-421 are of $7.17\pm0.66$~$M_{\oplus}$ for the inner planet and $16.42^{+1.06}_{-1.04}$~$M_{\oplus}$ for the outer one. For the younger system TOI-942, only the upper limits of planetary masses are constrained, and we assume them to be similar to the masses of the planets in the evolved TOI-421 system to facilitate comparison, namely 8.5 and 17~$M_{\oplus}$ for planets b and c, respectively. This choice, as well as the differences in the exact orbits, have a minor effect on our results (see detailed discussion in Paper I).}

\dik{The radii of {the} two outer planets coincide within the uncertainty, and are of $4.793^{+0.410}_{-0.351}$~$R_{\oplus}$ (TOI-942~c) and $5.09^{+0.16}_{0.15}$~$R_{\oplus}$ (TOI-421~c). In case of the inner planets, the values are $4.242^{+0.376}_{-0.313}$~$R_{\oplus}$ for TOI-942~b, and substantially smaller, $2.68^{+0.19}_{-0.18}$~$R_{\oplus}$, for TOI-421~b. In both systems, the radii of each planet changes a lot throughout the evolution, and therefore cannot be directly compared. The same holds for the XUV (X-ray+EUV) fluxes received by planets, as is seen from Figure~\ref{fig::stellar_models}. The present-day fluxes are about 50 times higher for the younger system.}

\section{Atmosphere evolution model}\label{sec::model}

In this study, we employed the planetary atmosphere evolution framework developed in \citet{kubyshkina2020mesa,kubyshkina2021mesa} to track the evolution of planets in the younger system TOI-942 and constrain the past of the TOI-421 system.

This framework is based on combining the thermal evolution of a planet performed with MESA models \citep[Modules for Experiments in Stellar Astrophysics,][]{paxton2011,paxton2013,paxton2018,paxton2019} and the realistic prescription of the atmospheric escape \citep{kubyshkina2018grid,kubyshkina2018approx} to track the evolution of planetary atmospheric parameters with time. It accounts for the heating of the planetary atmosphere by the star and allows for the inclusion of an arbitrary stellar evolution model. Here, we employ the stellar evolutionary tracks described in Section~\ref{sec::target_systems}.

The main difference between the evolution framework used in this study to that described in \citet{kubyshkina2020mesa,kubyshkina2021mesa} is the approach used to calculate the hydrodynamic atmospheric mass loss rates throughout the evolution. In \citet{kubyshkina2020mesa}, the atmospheric escape was calculated by employing the analytical approximation \citep[][]{kubyshkina2018approx} based on the large grid of the hydrodynamic 1D upper atmosphere models covering a wide range of planetary and stellar parameters. In the present study, we substitute it by the direct interpolation within the extended version of the grid of models presented in \citet{kubyshkina2021RN}. The extended grid covers the planets with masses of 1-109~$M_{\oplus}$ and radii of 1-10~$R_{\oplus}$ orbiting not further than the habitable zone of the host star, around stars between 0.4 and 1.3~$M_{\odot}$ and under different levels of stellar XUV irradiation. For more detail, see \citet{kubyshkina2018grid,kubyshkina2021RN}. For the present-day planetary and stellar parameters in TOI-421 and TOI-942 systems, this method predicts the atmospheric escape rates similar to the escape rates obtained in 3D models for both systems. 

\dik{To reduce computation time, the atmospheric mass-loss rates are not calculated at every time step but set instead constant for the specific time intervals. The duration of each specific interval is calculated on the basis of the current mass of the atmosphere and atmospheric mass-loss rate in a way that no more than 5\% of the whole atmosphere is lost within this time interval.}

\section{Can TOI-942 actually be the past of TOI-421 system?}\label{sec::evolution}

\subsection{TOI-942 b and c and their future}

To estimate the evolution of planets TOI-942 b and c, we have considered planets with their parameters, starting their evolution at disk dispersal with initial atmospheric mass fractions of 0.05-0.15$M_{\rm pl}$ for the inner planet, and of 0.05-0.4$M_{\rm pl}$ for the outer planet (with non-uniform steps increasing from 0.01 at lowest atmospheric mass fractions to 0.05 above $f_{at,0} = 0.15$). These ranges extend somewhat above the estimates we made for planets in TOI-942 systems in Section~\ref{sec::target_systems} for two reasons. First, we start the simulation at the disk dispersal time instead of at the current age of TOI-942 system to ensure that we can reproduce the two planets accounting for the effects of intense atmospheric escape at the beginning of the evolution (so, the planets are expected to lose some of their atmospheres before reaching the age of the system). And second, in the present work we employ the lower initial entropies (hence, luminosities/temperatures) of the planet cores compare to that used in \citet{kubyshkina2021mesa}, and therefore require the higher atmospheric mass fraction to reproduce the observed radii. 
%, 0.06, 0.07, 0.08, 0.1, 0.12, and
%, 0.06, 0.07, 0.08, 0.1, 0.12, 0.15, 0.2, 0.25, 0.3, 0.35, and

The effects of uncertain initial temperatures (luminosities) of planets are expected to be relevant for the first few tens of megayears \citep[see, e.g.,][]{owen2020entropy,kubyshkina2021mesa}, i.e., it might be still important in the case of TOI-942 system. {Therefore, we do not use the approach used in \citet{kubyshkina2021mesa} (a single mass-independent initial entropy of 8.5~${k_{\rm b}}/{\rm baryon}$ used for all planets)} and employ the initial entropies of 7.3 and 7.7~${k_{\rm b}}/{\rm baryon}$, as given by the entropy-mass relation suggested in \citet{malskyRogers2020}. To ensure that this choice has no effect on the conclusions we make, we also probe the higher entropy of 8.5~${k_{\rm b}}/{\rm baryon}$ and found that all the effects of different initial entropy/temperature vanish within $\sim 5$~Myr after the disk dispersal for planet b, and within $\sim$40-100~Myr for planet c depending on the specific planetary parameters.

We have also considered two different disk dispersal times of 5~Myr (the average time, \citealt{mamajek2009}) and 10~Myr. These two times allow reproducing the radius of the planet b with slightly different ranges of initial atmospheric mass fractions: $\geq0.07$~$M_{\rm pl}$ or $\geq0.05$~$M_{\rm pl}$ for 5 and 10~Myr, respectively. For planet c, the difference caused by the different disk dispersal times is smaller than can be accounted with the grid of initial atmospheric mass fractions considered here. In both cases, the radius of the planet c can be reproduced at the age of the system by considering the initial atmospheric mass fractions in the range of $\sim$0.08 - 0.15.

\begin{figure}
  \includegraphics[width=\hsize]{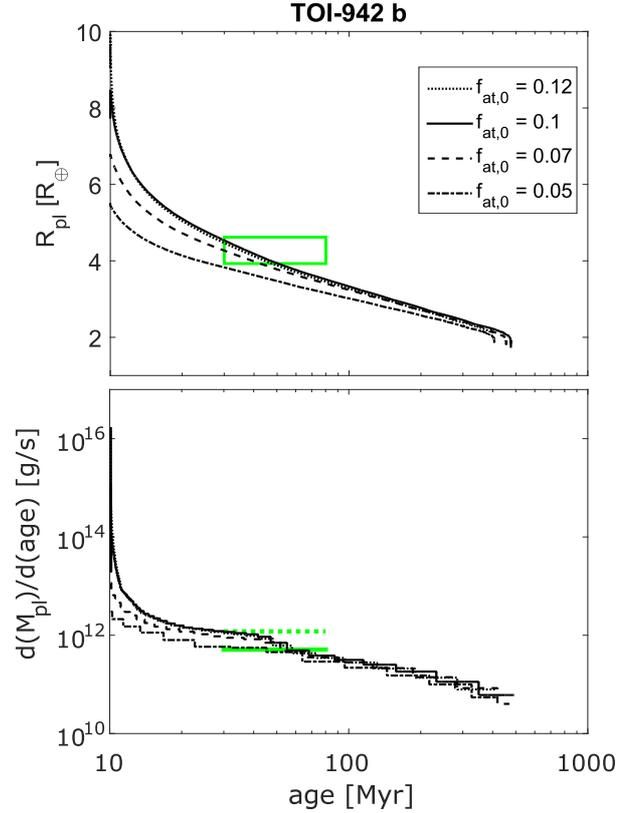}\\
  \caption{The evolutionary profiles for planetary radius (top) and atmospheric mass loss rate (bottom) of TOI-942~b assuming the ``fast rotator'' stellar evolutionary track described in Section~\ref{sec::target_systems}, initial entropy of the planet of 7.3 $\frac{k_{\rm b}}{\rm baryon}$ and the disk dispersal time of 10~Myr. Different types of line correspond to different initial atmospheric mass fractions of 0.05 (dashed-dotted), 0.07 (dashed), 0.1 (solid) and 0.12 (thin dotted line). The green rectangle in the top panel denotes the parameters of TOI-942~b from \citet{Carleo_toi942}, and the green lines in the bottom panel denote the atmospheric mass loss rates from 3D simulations (solid for the basic model and dotted for the split-wavelength one) at the age of the system.}\label{fig::evolution}
\end{figure}

In Figure~\ref{fig::evolution}, we show four evolutionary tracks for the inner planet TOI-942~b, assuming the disk dispersal time of 10~Myr and initial atmospheric mass fractions of 0.05, 0.07, 0.1, and 0.12~$M_{\rm pl}$. In the top panel, we show the radius evolution, where the radius and age of TOI-942~b known from observations are shown by the green rectangle. In the bottom panel, we show the corresponding atmospheric mass loss rates, where the green lines correspond to the values obtained with the 3D simulations. One can see, that the two tracks with the largest initial atmospheric mass fractions converge after a negligible time of $\sim0.12$~Myr (looking carefully, one can see the track with $f_{\rm at,0} = 0.12$ starts at $R_{\rm pl} = \sim10 R_{\oplus}$, and the track with $f_{\rm at,0} = 0.1$ at $\sim8.5 R_{\oplus}$). This effect is caused by the atmospheric mass-loss rate increasing with atmospheric mass fraction, and, hence, the radius of the planet \citep[see][and references therein]{kubyshkina2021mesa}. Therefore, the upper limit of the initial atmospheric mass fraction can only be constrained by employing planetary formation/atmospheric accretion models. Here we do not consider such a task.

The evolutionary tracks in Figure~\ref{fig::evolution} stop before 1~Gyr because all of the atmosphere is lost from the planet after about 500~Myr of evolution. The escape rates at the system's age ($50^{+30}_{-20}$~Myr), however, are similar to those predicted in 3D modeling (the green solid line in the bottom panel corresponds to the escape rate assuming the basic XUV model, and the dotted green line, assuming the split-wavelength XUV model) for any initial atmospheric mass fraction. Therefore, for TOI-942~b ($R_{\rm pl} \sim 4.24$) to have parameters similar to the TOI-421~b ($R_{\rm pl} \sim 2.68$) after Gyrs of evolution, the assumed planetary mass is insufficient.

Considering the mass of TOI-942~b at the upper limit of the observational constraint ($16$~$R_{\oplus}$) and the initial atmospheric mass fractions of 0.12-0.2 would allow to get similar radius to the radius of planet TOI-421~b at ages older than 6.3~Gyr (remnant atmospheric mass fraction would be of 0.02-0.04), but not a similar mass -- that would remain about twice higher than the actual mass of TOI-421~b. These models would also reproduce TOI-942~b only at the lower part of this $f_{\rm at,0}$ interval, 0.12-0.14. For the whole range {of $f_{\rm at,0}$},  the radii predicted at the age of 30-80~Myr are $\sim$5.1-5.7~$R_{\oplus}$. Thus, we can conclude that TOI-421~b could not be similar to TOI-942~b in the past.

\begin{figure}
    %\centering
    \includegraphics[width=\hsize]{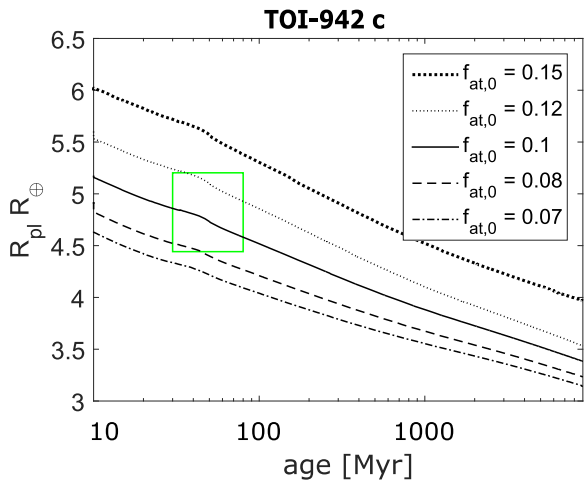}%alternative is rpl_age_print.eps
    \caption{\dik{The evolutionary profiles for planetary radius of TOI-942~c assuming the ``fast rotator'' stellar evolution track described in Section~\ref{sec::target_systems}, initial entropy of the planet of 7.7 $\frac{k_{\rm b}}{\rm baryon}$ and the disk dispersal time of 10~Myr. Different types of line correspond to different initial atmospheric mass fractions of 0.07 (dashed-dotted), 0.08 (dashed), 0.1 (solid), 0.12 (thin dotted), and 0.15 (thick dotted line). The green rectangle denotes the parameters of TOI-942~c from \citet{Carleo_toi942}.}}
    \label{fig::evolution_toi942c}
\end{figure}

Concerning the outer planet, the observed radius of TOI-942~c can be reproduced for the mass of the planet adopted in 3D simulations if assuming the initial atmospheric mass fraction of $\sim$0.08-0.14 (see Figure~\ref{fig::evolution_toi942c}). In this case, the planet is likely to keep a substantial part of its atmosphere throughout the evolution ($f_{\rm at}\approx$0.04-0.07 at 9~Gyr). The corresponding radius of the planet at the ages older than 6.3~Gyr is, however, $\sim 3.5$~$R_{\oplus}$, which is much smaller than the observed radius of TOI-421~b ($\sim 5.1$~$R_{\oplus}$): by about 30\% compared to the observational uncertainty of $\sim3\%$. Increasing the planetary mass, in this case, would not change the outcome significantly: at the upper mass limit, TOI-942~c would be reproduced with the similar range of the initial atmospheric mass fractions (0.07-0.12), and its radius would evolve to $\sim 4$~$R_{\oplus}$ at the age of TOI-421. 

\subsection{TOI-421 b and c and their past}

\dik{To obtain the results described above, we used a basic stellar model (as the source of bolometric and high-energy irradiation exposed at the planet) reproducing the present-day parameters of both TOI-942 and TOI-421, the ``fast rotator'' described in Section~\ref{sec::target_systems}. As we discussed in Paper I, this model describes the star rotating faster than average at young ages \citep{irwin2008}, which was not necessarily the case for TOI-421.} Therefore, to resolve the possible initial parameters of the planets in TOI-421 system, we consider {new stellar evolutionary models that reproduce well the present parameters of TOI-421,}
%the models assuming all the equivalent parameters (as they reproduce well the present parameters of the star), 
but {have} different rotation periods at 35~Myr compare to the 3.39 days assumed for the fast rotator (see Section~\ref{sec::target_systems}): 5 days (medium rotator for the given stellar mass \citealt{irwin2008}), 10 days (slow rotator), and 15 days (very slow rotator). %``Track 1'' is thus employed as a fast rotator.
%As the system TOI-942 is very young, the choice of any of these stellar models affect the range of initial atmospheric mass fractions allowing to reproduce the radii of planets b and c at the given age.
%To summarize, we consider $7\times2\times4 = 56$ individual evolutionary models for planet b and $11\times2\times4 = 88$ models for planet c. 
%Of them, 48 and 32, respectively, allow reproducing the radii of planets TOI-942~b and c known from the observations at the given age of the system. 
None of these {new} stellar evolution models let us reproduce the radii of planets in the TOI-421 system if assuming initial parameters similar to those of TOI-942 b and c. 

For there to be a chance of fitting both the mass and the radius of TOI-421~b, the initial mass of the planet can only be increased (which would allow keeping the primordial atmosphere long enough) with simultaneous increase of the initial atmospheric mass fraction, so this additional mass can be removed from the planet throughout the evolution. Thus, taking into account the present mass of TOI-421~b with the observational uncertainties, each potential initial planetary mass corresponds to a relatively narrow range of possible initial atmospheric mass fractions. To estimate the possible initial parameters of TOI-421~b, we run multiple evolutionary models employing the stellar evolution models described above and assuming different initial planetary masses and the corresponding intervals of initial atmospheric mass fractions outlined in Table~\ref{tab:toi421b_initials}. For each interval of $f_{\rm at,0}$, we take at least 3 values. We further interpolate the final planetary parameters ($R_{\rm pl,0}$ and $M_{\rm pl,0}$) at the age of TOI-421~b within these intervals to find for which exact $f_{\rm at,0}$ we can reproduce the radius of TOI-421~b for the specific planetary mass. The disk dispersal time we assume is 10~Myr.

\begin{table}
    \centering
    \begin{tabular}{c|l}
    \hline
      $M_{\rm pl,0} [M_{\oplus}]$   & $f_{\rm at,0} [M_{\rm pl}]$ \\
      \hline
       14.4  &  0.45 - 0.55\\
       15.0  &  0.48 - 0.57\\
       %15.2  &  0.49 - 0.57\\
       15.4  &  0.49 - 0.58\\
       16.0  &  0.51 - 0.59 \\
       17.0  &  0.53 - 0.62 \\
       18.0  &  0.56 - 0.64\\
       19.0  &  0.59 - 0.66\\
       20.0  &  0.61 - 0.67\\
       22.0  &  0.65 - 0.80\\
       24.0  &  0.67 - 0.75\\
       \hline
    \end{tabular}
    \caption{The initial masses and the corresponding ranges of initial atmospheric mass fractions used to model the evolution of TOI-421~b. To compare, the present day mass of TOI-421~b is $\sim 7.2$~$M_{\oplus}$.}
    \label{tab:toi421b_initials}
\end{table}

\begin{figure}
    \centering
    \includegraphics[width=\hsize]{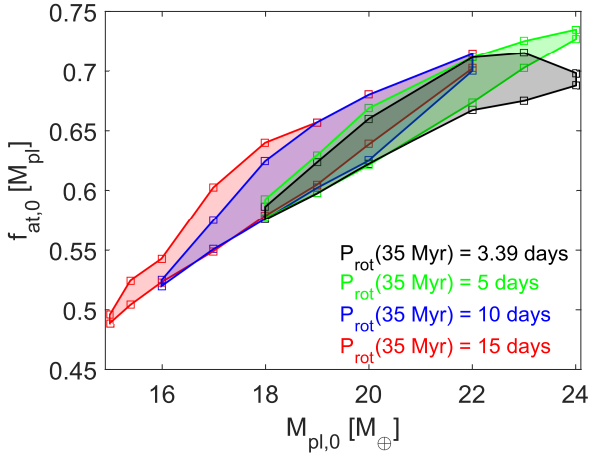}\\
    \caption{The initial parameters (initial atmospheric mass fraction against initial mass of the planet) of TOI-421~b constrained by our evolutionary model for different stellar rotation scenarios. The shaded areas correspond to the possible range of initial parameters, color-coded by different initial rotation periods, given at 35~Myr. The square markers denote the ranges resolved for the specific masses given in Table~\ref{tab:toi421b_initials}.}
    \label{fig::toi421b_initials}
\end{figure}

In Figure~\ref{fig::toi421b_initials}, we present our results (initial atmospheric mass fraction against initial mass allowing to reproduce the mass and the radius of TOI-942~b at the present time according to our evolution models) for different types of stellar rotation histories considered in this section. Even though the possible ranges of the initial atmospheric mass fraction outlined in Table~\ref{tab:toi421b_initials} for the specific initial planetary mass are already narrow, the actual parameter regions allowing to reproduce the radius and the mass of TOI-421~b at the age of the system are even narrower and are different for different rotation histories of the host star. Thus, naturally, for the faster rotating (hence, emitting higher XUV) star, the initial mass of the planet has to be on average larger compare to the slower rotator. In particular, for the fastest rotator, the initial mass of the planet has to be in the range of $\sim$17.5-24~$M_{\oplus}$, which corresponds to the overall range of initial atmospheric mass fractions of $\sim$0.57-0.71 (while the ranges for the specific masses remain very narrow). For {the} slowest rotator, these ranges are $\sim$15-22~$M_{\oplus}$ and $\sim$0.48-0.71. 

Assuming the same initial mass of the planet, TOI-421~b had to start with a smaller (maximum) initial atmospheric mass fraction if it orbits the fast rotator, compared to the slow one. This is because, near the faster rotator, the planet with given parameters has to stand stronger evaporation due to the higher irradiation, and more compact atmospheres were shown to be more stable \citep[see, e.g.,][]{kubyshkina2021mesa}.

Overall, the range of initial parameters allowing to fit the present-day parameters of TOI-421~b is broader, and the possible initial planetary masses are lower if assuming that TOI-421 was a slow rotator in its youth, compared to the case of the fast rotator. However, to make more robust conclusions one would need to employ a statistical framework to account for all the observational uncertainties (in the present study we only account for uncertainties on planetary mass and radius) and resolve the rotational history of the star in a probabilistic way, as was done in, e.g., \citet{kubyshkina2019a}. At the moment, our evolutionary framework is too computationally expensive for use in such an analysis (requiring an order of $10^5$-$10^6$ of single runs), so we leave this to future work.

%To clear up this possibility, we ran three additional models assuming the initial masses of 10.3, 12.0, and 14.4~$M_{\oplus}$ and respective initial atmospheric mass fractions of 0.3, 0.4, and 0.5. For two lower masses, the atmosphere of a planet is lost within 2 and 2.5~Gyr, respectively. \dik{The third simulation is yes running, if it fits or nearly fits, than TOI-421~b can be reproduced with a very narrow range of initial parameters. I'll have to double check with other rotators. Or it can't be fit at all, which is also funny.}

The above makes TOI-421~b a peculiar planet. First, it can be reproduced assuming only a narrow range of initial atmospheric mass fractions for the specific mass, and, second, such a high initial atmospheric mass fraction looks unlikely for a planet of (initial) 14-24~$M_{\oplus}$ mass at the orbital separation of 0.056~AU. We will come back to this discussion in Section~\ref{sec::conclusions}.

To fit TOI-421~c, one has to assume the initial atmospheric mass fraction of $\sim 0.30\pm0.07$. On the other hand, this can be done for essentially any type of stellar rotator, as stellar history does not affect the estimate on the initial atmospheric mass fraction of TOI-421~c much (as well as varying the initial mass of the planet within the range corresponding to the measured mass uncertainty of TOI-421~c). This is consistent with the estimates made in \citet{Carleo_toi421} with the simpler model (not accounting consistently for the thermal evolution of a planet), but with a more thorough probabilistic approach.

%We summarize here briefly the outcomes of this Section. First of all, the answer to the question placed in its title is most certainly negative. 

{To summarize, from the new set of evolutionary models presented in this section, we conclude that the} two planets {in TOI-421} %of the older system 
were likely born with more substantial envelopes than those 
%of the planets 
in the TOI-942 system. In the case of the inner planet, it also had to start with essentially higher mass, with most of it in the atmosphere. The planets in a young system TOI-942 will most likely end up after a few Gyrs of evolution on the different sides of the radius gap, with planet c keeping most of its atmosphere, and planet b losing all of it unless its true mass is close to the upper boundary of the existing constraint. In the latter case, planet b can preserve an atmosphere of a few percent of its mass.

%%%%%%%%%%%%%%%%%%%%%%%%%%%%%%%%%%%%
\section{Discussions and conclusions}\label{sec::conclusions}

\dik{In paper I of this series \citep{kubyshkina2021tois1}, we modeled atmospheric escape of the four planets in TOI-421 and TOI-942. Our 3D modeling approach allowed us to quantify how escape is affected by the stellar wind, which is particularly important at young ages, when the wind is believed to be stronger. We demonstrated that the wind can substantially affect the observational signatures of escape, but in the case of these systems, the effects on escape rate were not significant to alter the evolution of the planets. With that in mind, here, we employed our evolutionary framework (which neglects the presence of stellar winds) to investigate whether TOI-942 will one day become like TOI-421 (forward modeling) or, alternatively, whether TOI-942 is the past version of  TOI-421 (backward modeling). }

%evolution below
For that, we employed
%We have also considered the possible evolution scenarios of TOI-421 and TOI-942 systems, employing 
the evolution framework introduced in \citet{kubyshkina2021mesa}, with some adjustments. In the case of the young system TOI-942,  \dik{our forward modeling} mainly predicts the possible future of the system after Gyrs of evolution. Taking into account the rotation period of the host star, which is faster than average, we predict that the inner planet of the system will likely lose all of its atmosphere before the age of 1~Gyr, and will therefore cross the radius gap and become a super-Earth type planet. The outer planet, TOI-942~c, will likely keep some of its envelope and remain in the sub-Neptune group. Its predicted atmospheric mass fraction and the radius at the age of the TOI-421 system ($9.4^{+2.4}_{-3.1}$~Gyr) is of $\sim$0.045-0.095 and $\sim$3.5-4.0~$R_{\oplus}$ (compared to $5.09^{+0.16}_{-0.15}$~$R_{\oplus}$ of TOI-421~c), respectively. The TOI-942 planetary system cannot therefore be the past state of the TOI-421 system, and this remains true if we assume a slower rotating star.  \dik{In summary, the planetary system TOI-942 will presumably end up as one of the systems hosting two planets of relatively similar masses but lying on different sides of the radius gap, i.e., with a closer-in super-Earth and a further-out sub-Neptune planet. Examples of such systems are, e.g., HD~3167 \citep{christiansen2017} and Kepler-36 \citep{Vissapragada2020}. } %copying the paragraph from the intro

In the case of the old system TOI-421, the task  \dik{of performing our backward modeling} is less trivial. We know the present-day state of the system and would like to resolve its past. To achieve it, we consider a range of possible initial parameters of the planets in the system. We then run the forward evolution model and check a posteriori that models fit the observed parameters at the current age of the system. This process is further complicated by the unknown rotational history of TOI-421. To account for it, we have considered four different stellar evolution models assuming the  \dik{same internal stellar parameters,}
%same parameters as the \dik{fast rotator (reproducing the present-day parameters of TOI-942) described in Section~\ref{sec::target_systems}}, 
but using different rotation periods {at} 35~Myr: 3.39, 5, 10, and 15 days. This range of periods covers well the possible rotation scenarios for stars of mass similar to TOI-421.

Accounting for the above, we find that, to agree with the present state of the system, both TOI-421~b and c had to start their evolution with substantial envelopes. For planet c, our estimate is similar to the one made in \citet{Carleo_toi421}: its initial atmospheric mass fraction had to be about 0.3, and the initial mass had to be $\sim$15-35\% higher than the present-day mass. This estimate is only weakly dependent on the rotation history of the star. 
{For planet b,}
%In turn, the initial parameters of planet b were not previously constrained. In the present work, 
we estimate the overall range of possible initial masses (accounting for the various possible rotation histories of TOI-421) {between} 15 {and} 24~$M_{\oplus}$, and the corresponding range of initial atmospheric mass fractions of 0.48-0.71. The latter is the range 
for the whole mass interval. For each specific initial planetary mass, the range of possible initial atmospheric mass fractions is quite narrow, with a maximum absolute width of $\sim0.065$ (see Figure~\ref{fig::toi421b_initials}). These estimates are dependent on the rotation history of the star. Thus, the ranges of possible initial parameters are wider if assuming the slower rotator, and correspond to the lower average initial mass of the planet.

The initial parameters of TOI-421~b resolved in this study are peculiar, considering their compatibility with planetary formation models. More precisely, according to common atmospheric accretion models, a planet with such a substantial envelope could not form at the orbital distance of TOI-421~b. \citet{ikoma_hori2012} have demonstrated that the mass of the accreted atmosphere increases with increasing distance from the star and the mass of the core. Thus, according to their models, the planet with a core of about 7~$M_{\oplus}$ can accrete at $\sim0.05$~AU {an} atmosphere less than 0.01 of its mass. The analytical approximations given in the recent work by \citet{mordasini2020} predict the possible initial atmospheric mass fraction of TOI-421~b of $\sim0.01-0.02$, and the adaptation they present of the earlier work by \citet{lee_chiang2015} (adopting the same model parameters) suggests the possible range of 0.07-0.1. All these estimates are much below the range of 0.48-0.71 constrained by {our} atmospheric escape models. The same, though less extreme, applies to planet c.

The possible solution to this contradiction could be the formation of planets further out in the protoplanetary disk and the consequent migration during the disk phase. Relatively high eccentricities and orbital inclinations of the planets in TOI-421 system ($\sim 0.163$ and $85.7^{\circ}$, and $\sim0.152$ and $88.4^{\circ}$ for planets b and c, respectively) do not contradict this assumption \citep[see, e.g.,][]{izidoro2021}. Thus, according to the approximation based on \citet{lee_chiang2015}, TOI-421~b could be formed at $\sim$0.4-0.5~AU. This model lacks some of the relevant physical processes, and the approximation given in \citet{mordasini2020} does not allow to form TOI-421~b with 0.5-0.7 of its mass in the atmosphere within the snow line. However, \citet{michel2020} predicts that beyond the snow line, planets can be formed with a significant fraction of their mass being water ice (up to $\sim70\%$), which would be compatible with the initial parameters we constrained here. %Such scenario was previously considered for Kepler-444~b (Pezzotti et al., 2021).

The presence of the water vapor, in this case, could significantly reduce the heating efficiency \citep[][]{lopez2017}, and thus, the atmospheric mass-loss rates and the initial atmospheric mass fraction/planetary mass predicted for TOI-421~b. This possibility, however, requires more detailed consideration (including the self-consistent atmospheric chemistry modeling of TOI-942~b), and we leave it for future studies. \dik{Another assumption of our evolutionary model that can affect these results is the absence of  orbital migration after protoplanetary disk dispersal. {This post-disk migration} should be distinguished from the inwards migration during the disk phase discussed above, which affects the starting parameters of our simulations. Post-disk migration, on the contrary, affects the position of the planet (and thus its equilibrium temperature and amount of the XUV flux it receives) during the evolution simulation. %Thus, if the planets have migrated outwards, and had closer-in orbits in the past,  the initial XUV were larger and our estimate on the possible initial masses would change to the larger values.}
{Thus, if the planets have migrated outwards and had closer-in orbits in the past, the initial XUV must have been larger and our estimate on the possible initial masses would have to increase.}}% do you need this to be in red? if it is new text, it should be green, right?+

\dik{Finally, our models assume the absence of planetary magnetic fields.} The early paradigm considering the evolution of terrestrial planets has implied that the planetary magnetic field is necessary to protect planetary atmospheres and reduce the atmospheric mass loss \citep[see, e.g.,][and references therein]{dehant2007}. The later studies, however, show that this point of view is ambiguous. Thus, the effect of the magnetic field on the atmospheric escape can be considered as a result of the two concurring processes: reducing the escape by capturing the ionised atmospheric species within the closed magnetic field lines, and enhancing the escape of the atmospheric ions through the regions of the open magnetic lines (polar cusps, in the case of a dipole field) and the reconnection on the night-side \citep[see, e.g.,][]{khodachenko2015,sakai2018,carolan2021r}. Thus, for planets in the Solar System it was shown both in the observations \citep{gunell2018,ramstad2021} and by modeling \citep{sakai2018,egan2019} that the presence of a weak magnetic field can intensify  atmospheric escape. These results, however, should be taken with caution for young planets, and in particular those in the sub-Neptune range, because of the different atmospheric structure and the non-thermal mechanisms dominating the atmospheric mass loss in the Solar System, contrary to the planets considered in this study \citep[see, e.g.,][for the discussion]{scherf2021}. For hot Jupiters, \citet{khodachenko2015} predict a significant suppression of  escape for  intrinsic magnetic fields larger than 0.3~G. The model with the closest setup to the present study by \citet{carolan2021r} predicts, however, for the 0.7~$M_{\rm jup}$ planet experiencing XUV (thermally) driven atmospheric escape a small increase in the atmospheric mass-loss rate with increasing dipole field strength (about twice between 0 and 5~G). We therefore expect that, the possible effect from the planetary intrinsic magnetic field depends largely on strength and configuration of the  \dik{planetary and stellar magnetic} field, but, according to the numbers reported in the literature, {might} not affect our results dramatically. The lack of studies for close-in sub-Neptune-like planets, however, hold us from making final conclusions.

 \dik{To summarise our results, we demonstrated that the young planets b and c in the TOI-942 system are unlikely to follow the same evolutionary scenario as the planets b and c in the TOI-421 system, despite both b and c planets in the two systems being at similar close-in orbits and in similar mass ranges. With our forward modelling, we found that  the planetary system TOI-942 will presumably end up as one of the systems hosting two planets of relatively similar masses but lying on different sides of the radius gap, i.e., with a closer-in super-Earth and a further-out sub-Neptune planet. Examples of such systems are, e.g., HD~3167 \citep{christiansen2017} and Kepler-36 \citep{Vissapragada2020}. With our backward modeling, we found that planets in the TOI-421 system started their evolution with more massive envelopes compared to planets in the TOI-942 system, potentially suggesting their formation beyond the snow line. }

\section*{Acknowledgements}
This project has received funding from the European Research Council (ERC) under the European Union's Horizon 2020 research and innovation programme (grant agreement No 817540, ASTROFLOW). %The authors wish to acknowledge the SFI/HEA Irish Centre for High-End Computing (ICHEC) for the provision of computational facilities and support. This work used the BATS-R-US tools developed at the University of Michigan Center for Space Environment Modeling and made available through the NASA Community Coordinated Modeling Center.

\section*{Data Availability}
%TBC
The data underlying this article will be shared on reasonable
request to the corresponding author.
%The data underlying this article are available in Zenodo Repository, at  https://doi.org/...



%\bibliographystyle{mnras}


%%%%%%%%%%%%%%%%%%%%%%%%%%%%%%%%%%%%%%%%%%%%%%%%%%

%%%%%%%%%%%%%%%%% APPENDICES %%%%%%%%%%%%%%%%%%%%%

%\appendix
%

% Don't change these lines
\bsp    % typesetting comment
\label{lastpage}
\end{document}